\documentclass[conference,10pt]{IEEEtran}
\usepackage{cite,graphicx,psfrag,amsmath,amsfonts}
\usepackage{amsthm,balance,verbatim,multicol}
\usepackage{subfigure,multirow,booktabs,placeins,float}
\usepackage[dvipsnames]{xcolor}
\usepackage{steinmetz,relsize}
\usepackage{enumitem}

\DeclareMathOperator*{\argmax}{arg\,max}
\DeclareMathOperator*{\argmin}{arg\,min}

\begin{document}

\title{Preprint: Using RF-DNA Fingerprints To Classify OFDM Transmitters Under Rayleigh Fading Conditions}

\author{
  \IEEEauthorblockA{Mohamed Fadul, Donald Reising,T. Daniel Loveless, and Abdul Ofoli}\\ 
  \normalsize University of Tennessee at Chattanooga  \\
  \normalsize Chattanooga, TN 37403 USA \\
  \normalsize Email:\{lzw784, donald-reising, daniel-loveless, abdul-ofoli\}@utc.edu \\
}

\maketitle
\begin{abstract}
The Internet of Things (IoT) is a collection of Internet connected devices capable of interacting with the physical world and computer systems. It is estimated that the IoT will consist of approximately fifty billion devices by the year 2020. In addition to the sheer numbers, the need for IoT security is exacerbated by the fact that many of the edge devices employ weak to no encryption of the communication link. It has been estimated that almost 70\% of IoT devices use no form of encryption. Previous research has suggested the use of Specific Emitter Identification (SEI), a physical layer technique, as a means of augmenting bit-level security mechanisms such as encryption. The work presented here integrates a Nelder-Mead based approach for estimating the Rayleigh fading channel coefficients prior to the SEI approach known as RF-DNA fingerprinting. The performance of this estimator is assessed for degrading signal-to-noise ratio and compared with least square and minimum mean squared error  channel estimators. Additionally, this work presents classification results using RF-DNA fingerprints that were extracted from received signals that have undergone Rayleigh fading channel correction using Minimum Mean Squared Error (MMSE) equalization. This work also performs radio discrimination using RF-DNA fingerprints generated from the normalized magnitude-squared and phase response of Gabor coefficients as well as two classifiers. Discrimination of four 802.11a Wi-Fi radios achieves an average percent correct classification of 90\% or better for signal-to-noise ratios of 18 and 21 dB or greater using a Rayleigh fading channel comprised of two and five paths, respectively.
\end{abstract}
\section{Introduction}%
\label{sect:introduction}%
The Internet of Things (IoT) is based on two fundamental elements: 1) the Internet, and 2) semi-autonomous devices based upon inexpensive computing, networking, sensing, and actuating capabilities to sense and act within the physical world~\cite{DoD_IoT_2016}. The worldwide number of connected IoT devices is estimated to be 26.7 billion by the end of 2019 and be as high as 50 billion by 2020 and 75.4 billion by 2025~\cite{Gartner_2015,Juniper_2016,Statista_IoT_2019}. As highlighted by~\cite{Rawlinson_2014}, 70\% of IoT devices employ weak or no encryption; thus, as the number of deployed IoT devices continues to explode the development and integration of IoT security approaches becomes even more critical. The importance of IoT security is exacerbated as related vulnerabilities are exploited to conduct nefarious actions~\cite{Larsen_CNN_2017}.

Wireless networks are governed by the Open System Interconnect (OSI) model, which describes the  data units and services provided at each layer. Wireless networks traditionally implement security mechanisms within the OSI model's higher layers (e.g., Data-link layer), which ignores the Physical (PHY) layer. The PHY layer is the first layer exposed to malicious network attacks \cite{Collins2010}. Specific Emitter Identification (SEI) is one PHY layer approach that has been proposed for augmenting traditional digital security approaches such as those presented in \cite{DuanSpoof,LiSpoof}. SEI exploits physical characteristics, within the Radio Frequency (RF) waveform, to uniquely identify a wireless transmitter to prevent unauthorized network access \cite{ToonsKins95,UretenIEE99,DudczykSEI2,Jeffery_MobiCom_2007,JanaMobi08,BrikMobi08,Suski_IJESDF_2008,DanevIPSN09,Klein_ICC_2009,Liu_SEI_2009,Liu_SEI_2011,Kennedy_2010,Reising_IJESDF_2010,Reising_Dissertation,Williams_NSS_2010,Takahashi_CompApps_2010,TekbasIEE,Ellis_RadioSci,Soli_IEEE,CanadaSEI,azzouz,Reising_InfoSec_2015,Wheeler_ICNC_2017,Pan_2019}. A specific SEI approach, known as RF-Distinct Native Attributes (RF-DNA) fingerprinting, performs wireless transmitter discrimination by exploiting the distinct and unique coloration that is imparted to a portion of the waveform corresponding with a known sequence of bits (e.g., Wireless-Fidelity preamble). The distinct and unique waveform coloration is unintentionally created during waveform generation and transmission. This coloration is attributed to the characteristics, behaviors, and interactions that unintentionally exist within each transmitter's RF chain. Prior RF-DNA fingerprinting efforts have achieved wireless transmitter discrimination down to the serial number level; however, most assessed performance within an Additive White Gaussian Noise (AWGN) channel only \cite{Suski_IJESDF_2008,Klein_ICC_2009,Williams_NSS_2010,Reising_IJESDF_2010,Reising_InfoSec_2015,Wheeler_ICNC_2017}.
However, all wireless communication standards must contend with and implement processes for mitigating the effects of multipath fading.  Multipath is the reflection and/or scattering of the transmitted waveform, which results in multiple, differently attenuated, time delayed, and phase shifted copies being combined at the receiver~\cite{tse_2005}. \\
\indent Similar to \cite{Fadul_WCNC_2019,Fadul_thesis}, the focus of this work is to compensate for multipath fading at the waveform level. Selection of waveform level compensation is due to the
\vspace{-5px}
\begin{enumerate}[leftmargin=*]
    \item RF-DNA fingerprints being generated directly from the waveform. The goal is to compensate for multipath fading while preserving the RF-DNA exploited waveform features that facilitate radio specific discrimination.
    \item Elimination of additional, unnecessary processing (e.g., demodulation) for the case of unauthorized network devices/radios.
    \item Removal of the RF-DNA exploited features, that are mistaken as part of the channel impulse response, by constellation-based multipath estimation techniques. 
    Elimination of this constellation-based side effect would requires additional processing to preserve the features associated with each known/authorized radio prior to estimation of the channel response.
\end{enumerate}\vspace{-5px}
Based upon these considerations, the RF-DNA fingerprinting process has been modified to include waveform level multipath channel estimation and correction. This work differs from our work in \cite{Fadul_WCNC_2019,Fadul_thesis}, in that it:
\begin{enumerate}[leftmargin=*]
    \item Validates the waveform level Nelder-Mead (N-M) based estimator by comparing its performance with two additional channel estimation techniques: Least Square (LS) and Minimum Mean Squared Error (MMSE).
    \item Uses an MMSE equalizer instead of the Zero Forcing (ZF) technique. The MMSE equalizer innately considers the channel statistics; thus, making it better suited to multipath correction under degrading channel conditions (i.e., lower signal-to-noise ratio).
    \item Uses Rayleigh fading channels consisting of two, three, or five paths versus only two.
    \item Generates RF-DNA fingerprints using either the normalized magnitude-squared or phase angle of the complex Gabor Transform coefficients.
    \item Conducts classifier training using RF-DNA fingerprints extracted from waveforms that traversed a noisy, multipath fading channel as well as undergone channel estimation and correction versus waveforms that have \emph{only} traversed an AWGN channel.
    \item Performs classification using the neural network-based classifier known as Generalized Relevance Learning Vector Quantization-Improved (GRLVQI) in addition to Multiple Discriminant Analysis/Maximum Likelihood (MDA/ML) to facilitate comparative assessment.
\end{enumerate}
The remainder of this paper is arranged as follows: Sect.~\ref{sect:related_work} provides a summary related work, Sect.~\ref{sect:Background} provides an overview of the 802.11a Wireless-Fidelity (Wi-Fi) signal, signal collection and detection processes; Sect.~\ref{sect:Methodology} provides the Methodology that includes Multipath Channel Modeling, Time Offset Estimation, Least Square (LS) estimation, the Nelder-Mead based channel estimator, RF-DNA fingerprint generation and classification. RF-DNA fingerprint based SEI performance results are presented in Sect.~\ref{sect:Results} followed by the Conclusion in Sect.~\ref{sect:SumConc}.
\section{Related Work}%
\label{sect:related_work}%
This section provides a summary of the research efforts that have investigated performing SEI using signals collected within multipath fading environments \cite{Liu_SEI_2009,Liu_SEI_2011,Kennedy_2010,Reising_Dissertation,Fadul_WCNC_2019,Fadul_thesis,Pan_2019}. The work in \cite{Reising_Dissertation}, showed that multipath negatively affects SEI performance, because it distorts the exploited distinct and native waveform coloration. This is attributed to the very nature of multipath. For the purposes of SEI, it is desirable to remove or mitigate multipath channel effects while preserving the distinct and native waveform attributes that are exploited within the RF-DNA fingerprinting process.

The PHY layer work in~\cite{Liu_SEI_2009,Liu_SEI_2011} achieved SEI by exploiting constellation-based features associated with the transmitter's non-linear power amplifier coefficients. Unlike RF-DNA Fingerprinting, constellation-based SEI requires demodulation of the waveform before extracting features. The work in~\cite{Liu_SEI_2009,Liu_SEI_2011} modeled the power amplifiers using discriminatory features that remained fixed for each simulated transmission and across transmissions. However, it has been shown that a specific radio's RF front-end features do vary across its transmitted waveforms \cite{Wheeler_ICNC_2017}. In \cite{Liu_SEI_2011}, the multipath induced Inter-Symbol Interference (ISI) was suppressed using a linear approximation approach. The work used a multipath channel model with fixed channel coefficients; thus, multipath related distortion remained the same for all transmissions. This presents a point of potential bias within the multipath suppression as well as the subsequent radio discrimination.

The work in~\cite{Kennedy_2010} was the first to investigate waveform-based SEI in which the received waveforms were exposed to a multipath environment. The work employed an iterative approach that jointly estimated the multipath channel delay spreads and coefficients in conjunction with SEI. SEI was achieved by computing the residual power between the received waveform and each of the stored ``candidate'' waveforms. The received waveform was designated to have originated from the radio associated with the training waveform that resulted in the lowest, overall residual power. In~\cite{Kennedy_2010}, the signals were collected within an office environment in which multipath was stated to have been present; thus, the multipath channel characteristics (e.g., number of paths) were not specified. Lastly, the presented SEI approach did not perform Signal-to-Noise Ratio (SNR) based assessment.

The work in~\cite{Reising_Dissertation} presents the first SEI effort in which wireless radio discrimination was performed using RF-DNA fingerprints extracted from waveforms that underwent multipath fading. In~\cite{Reising_Dissertation} the RF-DNA fingerprints were generated from waveforms that propagated through a Rician fading channel with a single path. The RF-DNA fingerprints were generated directly from the received waveforms; thus, no channel estimation nor equalization was performed. The lack of multipath channel compensation resulted in the average percent correct classification performance dropping from $\sim$95\% to $\sim$45\% at an SNR$=$18~{dB}.

In~\cite{Pan_2019}, SEI is performed using the Hilbert-Huang Transform in conjunction with a deep learning technique known as a deep residual network. Under Rayleigh fading, the deep residual network achieved an average identification accuracy of 90\% or better at SNR$\geq$20~{dB}. As adopted from~\cite{Liu_SEI_2009}, the presented results are generated using simulated signals that have simulated power amplifier features applied to them to represent five unique emitters. Unlike the work presented here, the specifics of the Rayleigh fading channel (e.g., number of paths) is not stated nor does the work perform multipath channel estimation and correction.
\section{Background%
\label{sect:Background}}
\subsection{Signal of Interest%
\label{sect:Signal_of_Interest}}
The work presented here uses waveforms transmitted by radios that employ the IEEE 802.11a Wi-Fi communications standard \cite{802.11}. The selection of IEEE 802.11a Wi-Fi is attributed to the following factors: (i) the waveform is a result of Orthogonal Frequency Division Multiplexing (OFDM), (ii) the significant amount of SEI research using this wireless standard \cite{Jeffery_MobiCom_2007,Suski_IJESDF_2008,Liu_SEI_2009,Takahashi_CompApps_2010,Liu_SEI_2011,Reising_InfoSec_2015,Wheeler_ICNC_2017,Fadul_WCNC_2019,Fadul_thesis}, (iii) it is a designated IoT communications standard \cite{Porkodi_IntelComp_2014}, and (iv) many current and future wireless communication systems such as: 802.11ac, 802.11ad, 802.11ax, Long Term Evolution (LTE), and the Worldwide Interoperability for Microwave Access (WiMAX), are based upon an OFDM scheme \cite{Lajos_2011}. As in~\cite{Fadul_WCNC_2019,Fadul_thesis}, this work uses the 802.11a preamble for time-offset estimation, estimation of the multipath channel impulse response, as well as the region from which RF-DNA fingerprints are extracted to facilitate SEI.
\subsection{Signal Collection \& Detection%
\label{sect:collect}}
\begin{figure}[t!]
  \centering
  \includegraphics[width=0.5\textwidth]{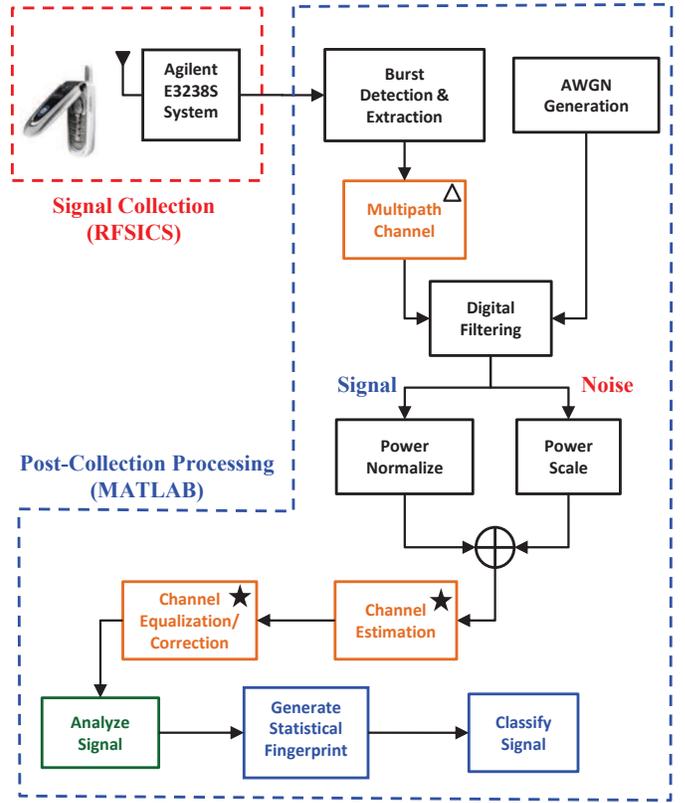}
    \caption{Signal collection and post-collection
processing adopted from~\cite{Reising_IJESDF_2010}. The multipath channel block ($\bigtriangleup$) were introduced in~\cite{Reising_Dissertation} while the channel estimation and correction blocks ($\star$) were added by \cite{Fadul_WCNC_2019}.}\vspace{-0.2in}
    \label{fig:the_process}
\end{figure}

Figure~\ref{fig:the_process} shows the overall process adopted by this work. This process is a modified version of the original RF-DNA fingerprinting process presented in~\cite{Reising_IJESDF_2010} with the inclusion of the multipath channel block from~\cite{Reising_Dissertation} and the channel estimation and equalization blocks added by~\cite{Fadul_WCNC_2019}. The set of IEEE 802.11a Wi-Fi signals used in this work are the same as those used to generate the RF-DNA results presented in~\cite{Reising_Dissertation,Reising_InfoSec_2015,Wheeler_ICNC_2017,Fadul_WCNC_2019}, which were collected from $N_{D}$$=$$4$ Cisco AIR-CB21G-A-K9 Wi-Fi cards operating in a peer-to-peer configuration and office environment. The signals were collected at a sampling rate of $f_{s}$$=$$95$~{MS$/$s} using an Agilent spectrum analyzer. Detection of individual transmissions, within a single collection record, is performed using the amplitude-based variance trajectory technique presented in~\cite{Klein_ICC_2009}. Following detection, every selected transmission underwent carrier frequency offset estimation and correction using the process presented in~\cite{Wheeler_ICNC_2017}. Lastly, the detected and CFO corrected signals were re-sampled from a rate of 23.75~{MHz} to 20~{MHz} to improve time synchronization performance~\cite{Fadul_WCNC_2019}.
\section{Methodology%
\label{sect:Methodology}}%
\subsection{Multipath Channel Modeling%
\label{sect:Multipath_Model}}
Multipath affects the performance of wireless communication networks significantly due to the destructive interference that results as each reflected signal is combined at the receiver. It is a major concern for indoor environments in which 802.11a Wi-Fi transceivers operate \cite{Ohara_Book_2005}. As in \cite{Ohara_Book_2005}, this work adopts a Rayleigh fading channel model to capture the statistical, time varying nature of an indoor multipath environment when one or more paths are present and the line-of-sight path does not. The Rayleigh fading channel model is the standard approach for predicting IEEE 802.11a Wi-Fi modulation performance within wireless multipath environments \cite{Ohara_Book_2005}.

Multipath is characterized by the time delay associated with each reflection path and is known as the delay spread \cite{Ohara_Book_2005}. The delay spread varies based upon the type of indoor environment. For example, the delay spread is below $50$~{ns} for home multipath environments, and around $100$~{ns} for office environments \cite{Ohara_Book_2005}. The multipath channel is modeled using a Tap Delay Line (TDL) in which each multipath component is represented by a single coefficient/gain and delay. For the case of a  Rayleigh fading channel, each path coefficient can be represented by a circularly symmetric complex Gaussian random variable given by,
\begin{equation}
	\alpha_{k} = A + jB,
	\label{eqn:complex_coeff}
\end{equation}
where $A$ and $B$ are zero mean independent and identically distributed Gaussian random variables with variance $\sigma^{2}$, and $k$ is the index of the multipath component \cite{Ohara_Book_2005}. The variance of $A$ and $B$ is given by,
\begin{equation}
	\sigma^{2} = \dfrac{1}{2}\left\{\left[1 - \exp\left(\dfrac{-T_{s}}{T_{\text{r}}}\right)\right]\exp\left(\dfrac{-kT_{s}}{T_{\text{r}}}\right)\right\},
	\label{eqn:variance}
\end{equation}
where $T_{\text{r}}$ is the Root-Mean-Squared (RMS) delay spread of the channel, and $T_{s}$ is the sampling period. For the channel models used to generate the results in Sect.~\ref{sect:Results}, the total power of the paths is normalized to ensure the same average received power, i.e.,
\begin{equation}
    \sum\limits_{k=1}^{L}{\sigma_{k}^{2}} = 1.
\end{equation}
where $\sigma_{k}$ is the variance of the $k^{\text{th}}$ path of the TDL. The statistics of the random variable, $\alpha$, are specified by it’s variance and magnitude given by the distribution,
\begin{equation}
	p(z) = \dfrac{z}{\sigma^{3}}\exp\left(\dfrac{-z^{2}}{2 \sigma^{2}}\right) , z \geq 0.
	\label{eqn:tab_magnitude}
\end{equation}
where $z$ is the magnitude of the coefficient $\alpha$. Equation \eqref{eqn:tab_magnitude} provides the probability density function of the magnitude of each path gain, $\alpha$, in the Rayleigh fading channel. After describing each multipath component as a complex random variable with Rayleigh distributed magnitude, the multipath channel comprised of $L$ paths is given as,
\begin{equation}
	h(t,\tau) = \sum\limits_{k=1}^{L}{\alpha_{k}\delta(t - \tau_{k}T_{s})},
	\label{eqn:tdl_eqn}
\end{equation}
where $\alpha_{k}$ is the complex gain of the $k^{\text{th}}$ path, with a variance given by \eqref{eqn:variance}, and $\tau_{k}$ is the delay spread of the $k^{\text{th}}$ path normalized by $T_{s}$ \cite{Ohara_Book_2005,Hijazi_TransVehicle_2009}. The path delays and variances used to generate the Rayleigh fading channel models, used in this work, are presented in Table~\ref{tab:multipath_channel_delays} and Table~\ref{tab:multipath_channel_variances}, respectively. 
\begin{table}[!t]
    \centering
    \caption{The delays used to generate the Rayleigh fading channel models comprised of $L$ paths.}
    \begin{tabular}{cccccc}
        \toprule
         {} & \multicolumn{5}{c}{Path Delays ($\tau_{k}^{2}$)} \\
         \cline{2-6}
         {$L$} & {$50$~ns} & {$100$~ns} & {$150$~ns} & {$200$~ns}  & {$250$~ns} \\
         \hline
         {$2$} & {$\surd$} & {$--$} & {$--$} & {$\surd$}  & {$--$} \\
         {$3$} & {$\surd$} & {$--$} & {$\surd$} & {$--$}  & {$\surd$} \\
         {$5$} & {$\surd$} & {$\surd$} & {$\surd$} & {$\surd$}  & {$\surd$} \\
         \bottomrule
    \end{tabular}
    \label{tab:multipath_channel_delays}
\end{table}
If the transmitted 802.11a signal is $x(t)$, then the received signal $y(t)$ filtered by a noisy, multipath channel is expressed as,
\begin{equation}
	r(t) = x(t) \ast h(t,\tau) + n(t),
	\label{eqn:rcvd_signal}
\end{equation}
where $\ast$ denotes convolution, and $n(t)$ is complex, white Gaussian noise with variance $\sigma_{n}^{2}$. The multipath channel in \eqref{eqn:tdl_eqn} is used to filter the IEEE 802.11a collected signals with SNR ranging from 9 to 30 dB in 3 dB steps.
\subsection{Time Offset Estimation%
\label{sect:Time_Offset}}%
Time offset estimation is the first step in estimating the channel impulse response. The estimated time offset is used to calculate the delay of the first path with respect to the receiver. The delays associated with any additional reflection paths are determined using the estimated value of the first path delay. The time offset estimation approach employed here was used in~\cite{Fadul_WCNC_2019} and is based upon the approach presented in~\cite{Wang_ATNAC_2003}. Time offset estimation is conducted using the normalized auto-correlation of the IEEE 802.11a preamble.
Given an OFDM signal, the output of a discrete-time multipath channel is expressed as,
\begin{equation}
	\resizebox{.91\columnwidth}{!}{$
	r(m) = \sum\limits_{k=0}^{L-1}{x(m - \theta - \tau_{k})h(k)\exp\left(\dfrac{j2\pi \varepsilon m}{N_{\text{c}}}\right) + n(m)},$}
	\label{eqn:rcvd_discrete}
\end{equation}
where $r(m)$ is the received signal, $m$ is the discrete-time index, $h(k)$ is the sampled complex channel impulse response, $\varepsilon$ is the carrier frequency offset, $n(m)$ is the complex white Gaussian noise, $N_{\text{c}}$ is the number of sub-carriers used to modulate symbols, $L$ is the length of the channel impulse response, and $\theta$ is the time offset to be estimated \cite{Wang_ATNAC_2003}.
\begin{table}[!t]
    \centering
    \caption{The normalized variance values used to generate Rayleigh fading channel models of $L$ paths.}
    \begin{tabular}{cccccc}
        \toprule
         {} & \multicolumn{5}{c}{Path Variances ($\sigma_{k}^{2}$)} \\
         \cline{2-6}
         {$L$} & {$\sigma_{1}$} & {$\sigma_{2}$} & {$\sigma_{3}$} & {$\sigma_{4}$}  & {$\sigma_{5}$} \\
         \hline
         {$2$} & {$0.8$} & {$--$} & {$--$} & {$0.2$}  & {$--$} \\
         {$3$} & {$0.8$} & {$--$} & {$0.13$} & {$--$}  & {0.07} \\
         {$5$} & {$0.865$} & {$0.117$} & {$0.016$} & {$0.002$}  & {$0.0003$} \\
         \bottomrule
    \end{tabular}
    \label{tab:multipath_channel_variances}
\end{table}

Estimation of the time offset is performed through the calculation of two normalized auto-correlation timing metrics using the preamble of the received signal. The first timing metric $M_{1}(\theta)$ is the normalized auto-correlation of the received signal with itself delayed by one Short Training Symbol (STS) duration ($0.8\mu$s) and is calculated using \eqref{eqn:timing_metric_1}. The result is a plateau that is the length of nine STS starting at the beginning of the first STS. The second timing metric $M_{2}(\theta)$, calculated using \eqref{eqn:timing_metric_2}, is the normalized auto-correlation of the received signal and a two STS delayed version of itself. The result of \eqref{eqn:timing_metric_2} is another plateau, but with a length of eight \cite{Wang_ATNAC_2003}.
\begin{equation}
	M_{1}(\theta) = \dfrac{\sum\limits_{m=0}^{N_{s}-1}{r(\theta + m)r^{*}(\theta + m + N_{s})}}{\sum\limits_{m=0}^{N_{s}-1}{|r(\theta + m)|^{2}}},
	\label{eqn:timing_metric_1}
\end{equation}
and
\begin{equation}
	M_{2}(\theta) = \dfrac{\sum\limits_{m=0}^{N_{s}-1}{r(\theta + m)r^{*}(\theta + m + 2N_{s})}}{\sum\limits_{m=0}^{N_{s}-1}{|r(\theta + m)|^{2}}},
	\label{eqn:timing_metric_2}
\end{equation}
For the two timing metrics $M_{1}(\theta)$ and $M_{2}(\theta)$, $N_{s}$ is the length of a single STS. The resulting two timing metrics are then used to determine the start of the received signal's ninth STS calculated by,
\begin{equation}
	\hat{\theta} = \argmax_{\theta}[M_{1}(\theta) - M_{2}(\theta)].
	\label{eqn:argmax}
\end{equation} 
If $\hat{\theta}$ is earlier than the true time, then part of the cyclic prefix of the \textit{current} symbol will be taken as data without ISI occurring. However, if $\hat{\theta}$ is later than the true time, then part of the cyclic prefix of the \textit{next} symbol will be taken as data and ISI will occur \cite{Wang_ATNAC_2003}.

\subsection{Channel Estimation%
\label{sect:CH_Estimation}}%
Following time offset estimation, the coefficients associated with the channel impulse response are estimated. In addition to the N-M estimator from \cite{Fadul_WCNC_2019}, this work uses two additional estimation approaches to validate the N-M estimation approach via comparative assessment. These two additional approaches are a Least Square (LS) and Minimum Mean Squared Error (MMSE) estimator. LS estimation is used in 802.11a Wi-Fi radios~\cite{Serra_Rapid_2004,Yuan_Inform_2008}. While the MMSE estimator accounts for the channel statistics at the expense of added computational complexity when compared to LS estimation \cite{Beek_1995,Edfors_1998}. Comparative assessment of the N-M, LS, and MMSE estimators is performed using: (i) accuracy of the estimated impulse response coefficients, and (ii) radio classification performance using RF-DNA fingerprints extracted from 802.11a Wi-Fi preambles that have undergone channel equalization using the coefficients estimated by each of the three techniques. The coefficient estimation accuracy assessment is conducted using ideal waveforms, average squared error for SNR$\in$[0, 30]~{dB}, and the results presented in~Sect.~\ref{sect:cmp_LS_NM}. The RF-DNA fingerprint classification assessment is performed using the collected waveforms (Sect.~\ref{sect:collect}), the MDA/ML classifier, average percent correct performance for SNR$\in$[9, 30]~{dB}, and the results presented in Sect.~\ref{sect:RF_DNA_Results}. The remainder of this section describes the LS, MMSE, and N-M estimation techniques.

\subsubsection{Least Square Channel Estimation%
\label{sect:LS_Estimation}}%
LS estimation of the channel's impulse response is conducted using the two 802.11a preamble Long Training Symbols (LTS). The frequency representation of the received LTS, $\mathbf{Y}$, is given by \eqref{eqn:H_Matrix}.
\begin{equation}
\resizebox{.89\hsize}{!}{$
\begin{bmatrix}
Y_{1} \\
Y_{2} \\
\vdots \\
Y_{N_{c}}
\end{bmatrix} = 
\begin{bmatrix}
H_{1} \\
H_{2} \\
\vdots \\
H_{N_{c}}
\end{bmatrix}
\begin{bmatrix}
X_{1} & 0 & \cdots & 0\\
0 & X_{2} &  \cdots & 0\\
\vdots & \vdots & \ddots & \vdots \\
0 & 0 & \cdots & X_{N_{c}}
\end{bmatrix}
+
\begin{bmatrix}
N_{1} \\
N_{2} \\
\vdots \\
N_{N_{c}}
\end{bmatrix}$,}
\label{eqn:H_Matrix}
\end{equation}
where $\mathbf{X}$ is the transmitted LTS' frequency response, $\mathbf{H}$ is the frequency response of the channel, and $\mathbf{N}$ is the frequency response of the complex white Gaussian noise. The LS estimator aims to minimize the cost function given by, 
\begin{equation}
	J(\hat{\mathbf{H}})=(\mathbf{Y}-\mathbf{H}\mathbf{X})^{H}(\mathbf{Y}-\mathbf{H}\mathbf{X}),
	\label{eqn:Cost_Function}
\end{equation}
where $\hat{\mathbf{H}}$ is the estimated frequency response of the channel, and $(\bullet)^{H}$ is the conjugate transpose of the matrix \cite{Serra_Rapid_2004}. The frequency response of the channel is determined by solving the cost function \eqref{eqn:Cost_Function} using the LS algorithm as follows,
\begin{equation}
	\hat{\mathbf{H}}_{L} = \mathbf{X}^{-1}\mathbf{Y}.
	\label{eqn:LS_H_Y}
\end{equation}
The impulse response of the channel is obtained by calculating the Inverse Discrete Fourier Transform (IDFT) of the LS estimate from \eqref{eqn:LS_H_Y}. The squared error, associated with the estimated values from \eqref{eqn:LS_H_Y}, is reduced by using both LTS' within the channel estimation process. This modified version of LS estimation is given by,
\begin{equation}
	\hat{\mathbf{H}}_{L}=\dfrac{1}{2}\mathbf{X}^{-1}(\mathbf{Y}_{1} + \mathbf{Y}_{2}),
	\label{eqn:LS_H_Y2}
\end{equation}
where $\mathbf{Y}_{1}$ and $\mathbf{Y}_{2}$ are the frequency responses of received preamble's first and second LTS, respectively~\cite{Yuan_Inform_2008}. In this case, the squared error corresponds to the variance of the channel noise.
\subsubsection{Minimum Mean Squared Error Channel Estimation%
\label{sect:MMSE_Estimation}}%
MMSE is a common channel estimation and equalization technique used in block- and comb-type OFDM systems. Based on the work in \cite{Beek_1995, Edfors_1998}, the MMSE is used to estimate the channel impulse response using the 802.11a preamble LTS as a pilot symbol. The use of the LTS eliminates the need for interpolation due to pilots being present within all of an LTS sub-carriers. 

Let the frequency domain representation of the estimated and actual transmitted LTS be $\hat{\mathbf{X}}$ and $\mathbf{X}$, respectively. The MMSE minimizes the error given by,
\begin{equation}
	\mathtt{e} = E[\hat{\mathbf{X}}-\mathbf{X}]^2,
	\label{eqn:MMSE_Error}
\end{equation}
where $E[\bullet]$ is the expectation. The MMSE estimator is implmented using the Linear MMSE (LMMSE) channel estimator in \cite{Edfors_1998}, which is given by,
\begin{equation}
	\hat{h}_{M}=\mathbf{R}_{hh}\left[\mathbf{R}_{hh}+\sigma_{n}^2(\mathbf{X}\mathbf{X}^H)^{-1}\right]^{-1}\hat{h}_{L},
	\label{eqn:MMSE}
\end{equation}
where $\hat{h}_{L}$ is the IDFT of the LS frequency response estimated by \eqref{eqn:LS_H_Y2}, $\mathbf{R}_{hh}$ is the auto-correlation matrix of the channel given by,
\begin{equation}
	\mathbf{R}_{hh}=E[hh^H],
	\label{eqn:channel_correlation}
\end{equation}
and $\sigma_{n}^2$ is the variance of the AWGN noise.
\subsubsection{Nelder-Mead Based Channel Estimation%
\label{sect:NM_Estimation}}
The Nelder-Mead (N-M) simplex algorithm is a computationally compact direct search method used in unconstrained optimization problems to achieve the minimum of a function \cite{Nelder_CompJrnl_1965}. The N-M algorithm is widely used in search and optimization problems due to its robustness and computational efficiency \cite{Wouk_Book_1987}. First an explanation of the N-M simplex algorithm is presented based upon the work in~\cite{Nelder_CompJrnl_1965,lagarias_Optim_2006}. This is followed by a description of how the N-M simplex algorithm is employed in the estimation of the multipath channel coefficients. 

The N-M simplex algorithm attempts to find the minimum of a $d$-variable non-linear function by using only function values \cite{Nelder_CompJrnl_1965,lagarias_Optim_2006}. This eliminates the need for computing derivatives, which aids in improving the computational efficiency of the algorithm \cite{lagarias_Optim_2006}. The problem to be solved by the N-M simplex algorithm is defined by,
\begin{equation}
	\operatorname*{minimize}_{x \in R^{d}}f(x).
	\label{eqn:minmize}
\end{equation} 
The N-M simplex algorithm requires definition of four parameters: reflection coefficient $(\rho)$, contraction $(\gamma)$, expansion $(\chi)$, and shrinkage $(\varphi)$. These four parameters should satisfy, 
\begin{equation}
	\rho > 0, \chi > 1, \chi > \rho, 0 < \gamma < 1, 0 < \varphi < 1.
	\label{eqn:NM_constraints}
\end{equation}
The N-M algorithm performs a sequence of iterations, $k$$\geq$$0$, to solve for the function's minimum. Each iteration begins with a simplex determined by $d+1$ vertices where $d$ is the number of the variables in the function to be minimized. For example, a $d+1$$=$$3$ simplex is a triangle on a plane corresponding to a function of $d$$=$$2$ variables. At the start of each iteration $k$, the simplex vertices are ordered based upon their corresponding function value,
\begin{equation}
    f(x_{1}) \leq f(x_{2}) \leq \cdots \leq f(x_{d+1}),
\end{equation}. 
where $x_{i}$$\in$$\mathbb{R}^{d}$ and $1$$\leq$$i$$\leq$$d+1$. Next, four operations of: reflection, expansion, contraction, and shrinkage are applied sequentially to generate one or a set of points. The condition associated with each operation is detailed in~\cite{lagarias_Optim_2006}, but is essentially the comparison of the function value corresponding to the calculated point, or set of points, to the best point $(x_{1})$ and the worst point $(x_{d+1})$ of the current simplex. The new calculated point is used to replace the worst point if it satisfies one of the conditions detailed in ~\cite{lagarias_Optim_2006,Nelder_CompJrnl_1965} and associated with the reflection, expansion, or contraction operations. If none of the points calculated using: reflection, expansion, nor contraction satisfy certain conditions, then a new set of simplex vertices are generated using the shrink operation. These new set of vertices are given by,
\begin{equation}
	v_{i}=x_{i}+\varphi(x_{i}-x_{1})
	\label{eqn:Shrinkage}
\end{equation} 
where $2$$\leq$$i$$\leq$$d+1$, and the new simplex for the next iteration is $(x_{1}, v_{2}, \cdots v_{d+1})$~\cite{lagarias_Optim_2006}. The algorithm is terminated when the function value, at the vertices of the $k^{\text{th}}$ simplex/iteration, satisfies a specific or set of conditions. In~\cite{Fadul_WCNC_2019} and here, two stopping conditions were used to terminate the iterative search process. The first stopping condition is based upon the function value at the simplex vertices and is given by,
\begin{equation}
	\dfrac{1}{d}\sum\limits_{i=1}^{d+1}{[f(x_{i}) - \bar{f}]^{2} < \epsilon_{1}},
	\label{eqn:condition_one}
\end{equation}
where $\bar{f}$  is the mean of the function values at all the vertices, and $\epsilon_{1}$ is a tolerance based on the function values. The second stopping condition, proposed in~\cite{Parkinson_NumMethods_1972}, relies upon the vertices themselves and is given by,
\begin{equation}
	\dfrac{1}{d}\sum\limits_{i=1}^{d}{\left\|x_{i}^{k} - x_{i}^{k+1}\right\|^{2} < \epsilon_{2}},
	\label{eqn:condition_two}
\end{equation}
where $\|\bullet\|$ is the $l_{2}$-norm. If the $l_{2}$-norm of the point $x_{i}$ between two successive iteration is less than the tolerance $\epsilon_{2}$, then the algorithm is terminated~\cite{Wouk_Book_1987}. These two conditions are checked at the end of each iteration.

Now that the operation of the N-M simplex algorithm has been explained, the remainder of this section explains its use in estimating the coefficients of the multipath channel's impulse response. Assuming time and frequency offset correction has been performed on the received signal given by \eqref{eqn:rcvd_discrete}, then the function to be minimized by the N-M simplex algorithm is,
\begin{equation}
	f(h) = \mathlarger{\sum}_{k \in m}{\bigg|r(m) - \sum\limits_{k=0}^{L-1}{x(m - \tau_{k})h_{k}\bigg|^{2}}},
	\label{eqn:to_be_minimized}
\end{equation}
which is a squared error function between the received signal $r(m)$ and the transmitted signal $x(m)$. In \eqref{eqn:to_be_minimized}, $h_{k}$ and $\tau_{k}$ are the multipath coefficient and delay corresponding to the $k^{\text{th}}$ reflection path. One limitation of the N-M simplex algorithm is that it cannot solve minimization problems for complex-valued functions \cite{lagarias_Optim_2006,Nelder_CompJrnl_1965}. In \eqref{eqn:to_be_minimized}, both the received signal $r(m)$ and ``candidate'' signal $x(m)$ are complex valued; thus, the N-M algorithm cannot be directly applied to it. This technicality is circumvented by expanding \eqref{eqn:to_be_minimized} and grouping the signals by their real and imaginary components. The result is two new squared error functions that are only real valued; thus, allowing the use of the N-M simplex algorithm. These two new squared error functions are:
\begin{eqnarray}\nonumber
	C_{1} = \sum_{m \in T}{\bigg|\Re\{r(m)\} - \sum\limits_{k=1}^{L}{[h_{k,r,1} \times  \Re\{x(m - \tau_{k})\}]}}\\ - [h_{k,i,1} \times  \Im\{x(m - \tau_{k})\}]\bigg|^{2},
	\label{eqn:expand_real}
\end{eqnarray}
and
\begin{eqnarray}\nonumber
	C_{2} = \sum_{m \in T}{\bigg|\Im\{r(m)\} - \sum\limits_{k=1}^{L}{[h_{k,r,2} \times \Im\{x(m - \tau_{k})\}]}}\\  - [h_{k,i,2} \times \Re\{x(m - \tau_{k})\}]\bigg|^{2},
	\label{eqn:expand_imag}
\end{eqnarray}
where $\Re\{\bullet\}$ represents the real parts, $\Im\{\bullet\}$ represents the imaginary parts, and the real and imaginary parts of the $k^{\text{th}}$ channel coefficient are [$h_{k,r,1}$, $h_{k,r,2}$] and [$h_{k,i,1}$, $h_{k,i,2}$], respectively. The N-M simplex algorithm is used in the minimization of \eqref{eqn:expand_real} and \eqref{eqn:expand_imag} to determine the optimal values of the corresponding coefficients. In an effort to reduce the estimation error, the average of the optimal channel coefficients are used in channel equalization~\cite{Fadul_WCNC_2019}. The averaged coefficients are calculated by,
\begin{equation}
	\bar{h}_{k,r} = \dfrac{h_{k,r,1} + h_{k,r,2}}{2},
	\label{eqn:real_avg}
\end{equation}
and 
\begin{equation}
	\bar{h}_{k,i} = \dfrac{h_{k,i,1} + h_{k,i,2}}{2}.
	\label{eqn:imag_avg}
\end{equation}
The averaged coefficients $\bar{h}_{k,r}$ and $\bar{h}_{k,i}$ represent the the final estimate of the real and imaginary parts of the channel coefficients for a given ``candidate'' preamble. In this work, a set of $N_{p}$$=$$20$ ``candidate'' preambles are used to represent $x(m)$ in \eqref{eqn:expand_real} and \eqref{eqn:expand_imag}. These candidate preambles were randomly selected from the $N_{B}$$=$$2,000$ total signals collected for each of the $N_{D}$$=$$4$ Wi-Fi radios; thus, each radio is represented by $N_{p}$$/$$4$$=$$5$ preambles. Due to the use of the candidate set of preambles, a total of $N_{p}$ channel estimates are obtained for a given received signal $r(m)$. Similar to~\cite{Kennedy_2010}, the ``best'' channel estimate is selected using residual power, which is calculated by,
\begin{equation}
	\hat{h}(m) = \argmin_{c}\left\{\sum_{m}{\big|r(m) - \hat{h}_{c}(m) \ast x_{c}(m)\big|^{2}}\right\},
	\label{eqn:best_estimate}
\end{equation}
where $1$$\leq$$c$$\leq$$N_{p}$, and $\hat{h}_{c}(m)$ is the estimated channel associated with candidate preamble $x_{c}(m)$~\cite{Fadul_WCNC_2019}. The set of the channel coefficients $\hat{h}_{c}(m)$ associated with the minimum residual power are selected as the ``best'' estimate of the channel and is designated as $\hat{h}_{B}(m)$.
\subsection{Channel Equalization%
\label{sect:CH_EQ}}%
Following estimation of the channel impulse response using either LS, MMSE, or N-M estimation, multipath channel effects are compensated using channel equalization. In~\cite{Fadul_WCNC_2019,Fadul_thesis}, channel equalization was implemented through the use of a filter with coefficients equal to the inverse of the estimated channel coefficients obtained via the LS or N-M approach. This equalization approach is known as Zero Forcing (ZF) equalization. For ZF equalization, the transmitted signal is estimated by,
\begin{equation}
    \hat{x}=\dfrac{1}{N_{K}}\sum\limits_{k=0}^{N_{K}-1}\dfrac{R(k)}{\hat{H}_{B}(k)}\exp{\left(j\dfrac{2\pi k m}{N_{K}}\right)},
    \label{eqn:Div_equalizer}
\end{equation}
where $0$$\leq$$m$$\leq$$N_{K}$, $N_{K}$ is the number of points comprising the  Discrete Fourier Transform (DFT), $R(k)$ is the received signal's DFT, and $\hat{H}_{B}(k)$ is the DFT of the ``best'' estimated channel, $\hat{h}_{B}(m)$. 

This work adds an MMSE equalizer, because it is more robust to degrading SNR due to integration of the channel statistics within its calculation of the transmitted signal. Thus, the MMSE channel equalizer should perform better than that of the ZF equalizer, which translates to better RF-DNA fingerprint classification performance as shown in Sect.~\ref{sect:results_class_trad_vs_mmse}. The goal of the MMSE equalizer is to minimize the squared error between the transmitted signal estimate $\hat{x}(m)$ and the original signal $x(m)$ by solving,
\begin{equation}
	\hat{x}(m) = \argmin_{\hat{x}(m)}E\left[(x(m) - \hat{x}(m))^{2}\right],
	\label{eqn:argmin_mmse}
\end{equation} 
where $E[\bullet]$ is the expected value. Inclusion of the channel statistics requires knowledge of either: (i) the power of the channel noise and transmitted signal or (ii) the channel SNR. If one of these two cases are known, then the MMSE equalizer can be used to recover the transmitted signal by calculating,
\begin{equation}
	\hat{x}_{M} = \mathbf{A}^{H}\left(\mathbf{A} \mathbf{A}^{H} + \gamma^{-1}\mathbf{I}_{A}\right)^{-1}r
	\label{eqn:mmse}
\end{equation} 
where $\gamma$ is the SNR, $\mathbf{A}$ is a square matrix representing the channel impulse response, $\mathbf{I}_{A}$ is an identity matrix of the same dimension as $\mathbf{A}$, and $r$ is the received signal~\cite{Rugini_Letter_2005}. In this work $\mathbf{A}$ is a diagonal matrix, because the Doppler shift is assumed to be zero. 
\subsection{RF-DNA Fingerprints generation%
\label{sect:RF_DNA_Generation}}%
This work adopts the same time-frequency (T-F) based RF-DNA fingerprint generation approach presented in \cite{Reising_Dissertation}. Prior works that have used Gabor Transform (GT)-based RF-DNA fingerprints, extracted the fingerprints from the normalized magnitude-squared GT coefficients. This work introduces the use of RF-DNA fingerprints generated from the phase of the Gabor coefficients. The Gabor coefficients are given by,
\begin{equation}
 G_{\eta k} = \sum_{m=1}^{MN_\Delta} s(m)W^{*}(m -\eta N_\Delta)\exp\left(-j\dfrac{2\pi km}{K_G}\right),
 \label{eqn:GaborExpansion}
\end{equation}
where $G_{\eta k}$ are the Gabor coefficients, $s(m)$$=$$s(m+lMN_\Delta)$ is a periodic input signal, $W(m)$$=$$W(m+lMN_\Delta)$ is a periodic analysis window, $N_\Delta$ is the number of shifted samples, $\eta$$=$$1,2,\dots,M$  for $M$ total shifts, and $k$$=$$0,1,\dots,K_G-1$ for $K_G$$\geq$$N_{\Delta}$ and $mod(MN_{\Delta},K_G)$$=$$0$ is satisfied. Gabor-based RF-DNA fingerprints are extracted from either the normalized magnitude-squared, $|G_{\eta k}|^2$, or the phase, $\phase{G_{\eta k}}$, of the Gabor coefficients calculated with $M$$=$$186$, $K_G$$=$$186$, and $N_\Delta$$=$$1$. The normalized magnitude-squared and phase responses each represent a T-F surface. For a given response, the T-F surface is subdivided into $N_{R}$ 2-D patches. Each patch contains $N_{T}$$\times$$N_{F}$ values, where $N_{T}$$ = $$12$ and $N_{F}$$ = $$10$ represents the length of the patch along the time and frequency dimension, respectively. Each patch is reshaped into a 1-D vector for feature calculation. The features used are: standard deviation, variance, skewness, and kurtosis. These features are also calculated for the entire T-F surface. The resulting RF-DNA fingerprints are comprised of $N_{f}$$=$$363$ features.

\subsection{RF-DNA Fingerprint Classification%
\label{sect:RF_DNA_Classification}}%
\subsubsection{Multiple Discriminant Analysis/Maximum Likelihood (MDA/ML)}
The Fisher-based MDA/ML classifier is adopted due to its simplicity as well as successful implementation and demonstration in previous RF-DNA fingerprinting efforts \cite{Reising_Dissertation,Wheeler_ICNC_2017}. MDA facilitates feature selection by linearly projecting the $N_{f}$$=$$363$ dimensional fingerprints into a $N_{D} - 1$ dimensional subspace that maximizes radio separability while simultaneously reducing within radio variance~\cite{dhsPC}. Following projection, a multivariate normal distribution is ``fit'' to the projected fingerprints of each radio to facilitate ML classification. Based upon Bayesian Decision Theory, an ``unknown'' projected RF-DNA fingerprint is estimated to have originated from the radio associated with the distribution that results the largest likelihood value. In Sect.~\ref{sect:RF_DNA_Results}, the percent correct classification is calculated based upon the number of times the classifier correctly assigns an ``unknown'' projected RF-DNA fingerprint over all trials.

\subsubsection{Generalized Learning Vector Quantization-Improved (GRLVQI)} RF-DNA fingerprint classification performance is also assessed using GRLVQI, an artificial neural network based classifier. The selection of GRLVQI is due to: (i) prior demonstrated RF-DNA fingerprinting success \cite{Reising_Dissertation,Reising_InfoSec_2015}, (ii) no required knowledge nor assumption of the input data distribution, (iii) inherent feature selection during classifier training, and (iv) it is well-suited to cases where the number of inputs vary across classes or the inputs represent noisy or inconsistent data. The third is of particular interest here. RF-DNA fingerprint performance is assessed under degrading SNR, which satisfies the noisy data case. Inconsistent data is attributed to the degrading SNR and multipath effects that remain after estimation and correction. GRLVQI uses a set of prototype vectors to define the boundary of each class'/radio's classification region. During training, these prototype vectors are ``repositioned'' within the $N_{f}$-dimensional space with the goal of minimizing the Bayes risk. A detailed explanation of the GRLVQI training process is presented in \cite{Mendenhall_TNN08}. Following training, an unknown RF-DNA fingerprint is said to have originated from the radio whose assigned prototype vector resulted in the minimum Euclidean distance. Comparative assessment of the GRLVQI and MDA/ML classifiers is performed using percent correct classification and presented in Sect.~\ref{sect:results_MDAML_GRLVQI}.

\section{Results%
\label{sect:Results}}%
For the results presented here, Rayleigh fading multipath channel models are implemented to simulate indoor environments comprised of $L$$=$$[2, 3, 5]$ paths using the delays and variances presented in Table~\ref{tab:multipath_channel_delays} and Table~\ref{tab:multipath_channel_variances}, respectively. 
\subsection{Channel Estimation: A Performance Comparison%
\label{sect:cmp_LS_NM}}%
\begin{figure}[!b]
	\centering
	\subfigure[$L$$=$$2$ Rayleigh fading paths.]{\label{fig:error_results_a}
	\includegraphics[width=0.75\columnwidth]{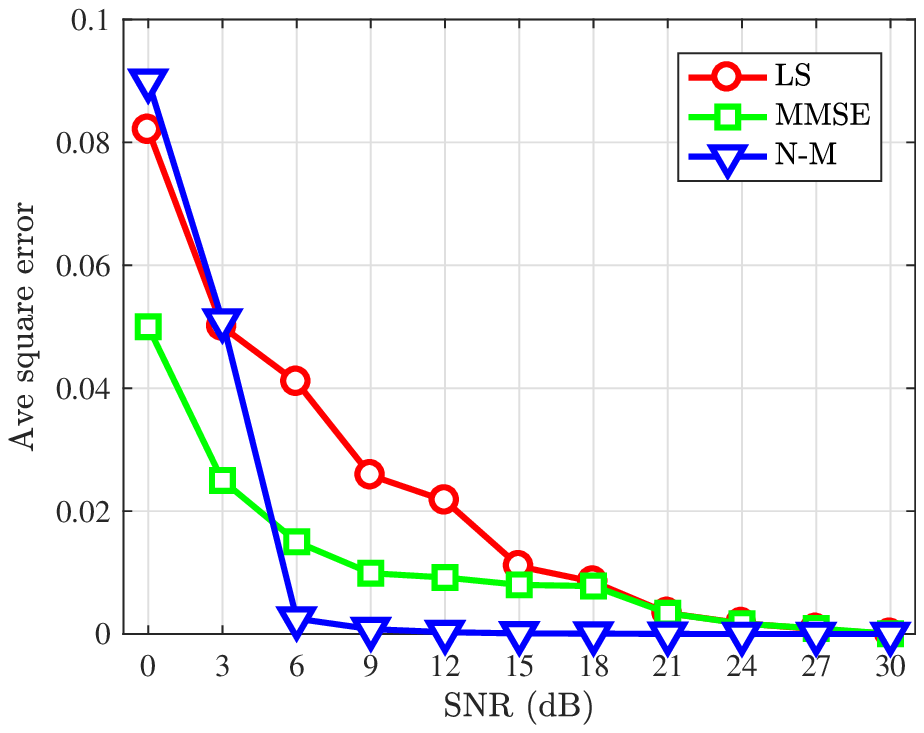}}
	\subfigure[$L$$=$$5$ Rayleigh fading paths.]{\label{fig:error_results_c}
	\includegraphics[width=0.75\columnwidth]{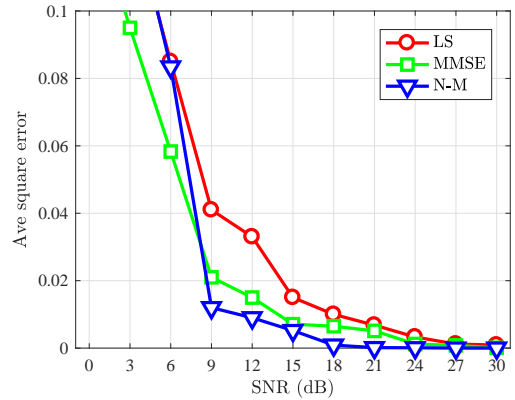}}
	\caption[]{The average squared error of the LS, MMSE, and N-M estimators at $SNR$$\in$$[0, 30]$~dB.}
	\vspace{-5mm}\label{fig:error_results}	
\end{figure}

Comparative assessment of the LS, MMSE, and N-M estimators is conducted using Rayleigh fading channel models comprised of $L$$=$$[2, 5]$ paths and expressed by \eqref{eqn:tdl_eqn}. For each of the $N_{D}$$=$$4$ Wi-Fi radios, a total of 1,000 preambles are randomly selected from their corresponding data set of $N_{B}$$=$$2,000$. A unique Rayleigh fading channel is generated and convolved with every selected preamble of the four Wi-Fi radios for the $L$$=$$2$ and $L$$=$$5$ cases. This equates to 4,000 unique Rayleigh fading channels for a given path length, $L$. Following application of a multipath channel, like-filtered AWGN is generated, scaled to achieve a specific SNR, and added to the selected preamble. This process is repeated for all preambles at SNR$\in$[0, 30]~dB in 3~dB increments. Additionally, Monte Carlo analysis is performed through the use of $N_{z}$$=$$10$ AWGN realizations per SNR; thus, for the results presented in Fig.~\ref{fig:error_results} a total of $4$$\times$$1,000$$\times$$10$$=$$40,000$ channel estimates are obtained at every SNR by each of the estimation approaches. Comparative assessment is conducted through the calculation of the squared error measure given by,
\begin{equation}
	\epsilon = \mathlarger{\sum}_{m \in L}{\bigg|h(m) - \hat{h}(m)\bigg|^{2}},
	\label{eqn:square_error_measure}
\end{equation}
where $h(m)$ is the actual channel coefficients and $\hat{h}(m)$ is the estimated channel coefficients determined by either the LS, MMSE, or N-M estimator. The squared error provides a measure of how close the estimated channel coefficients are to those of the actual channel.

For each of the three assessed estimators, the average squared error is calculated across all 1,000 preambles, $N_{D}$$=$$4$ radios, and noise realizations at each SNR and is presented in Fig.~\ref{fig:error_results}. For the case of $L$$=$$2$ paths, Fig.~\ref{fig:error_results_a}, the LS and MMSE estimators are outperformed by the N-M estimator for all SNR$\geq$6~{dB}. However, for SNR$<$6~dB, the MMSE estimator results in a more accurate estimated of the multipath channel coefficients. Figure~\ref{fig:error_results_c} presents the squared error results for the LS, MMSE, and N-M estimation performance when the fading channel consists of $L$$=$$5$ paths. The N-M estimator results in the smallest squared error for SNR$\geq$9~{dB}. However, the MMSE estimator results in the smaller squared error at SNR$\leq$6~{dB}. Effective demodulation of 802.11a Wi-Fi signals typically requires SNR$\geq$10~dB, which makes the N-M estimator the ``best'' option for performing channel estimation within the RF-DNA fingerprinting process~\cite{802.11}. The MMSE estimator's performance improvement for lower SNR values is expected as it is the only one that accounts for the channel statistics within the estimation process.
\begin{table*}[!t]
    \centering
    \caption{Average percent correct classification performance using the MDA/ML classifier and the channel coefficients are estimated using either N-M, MMSE (\textit{ITALICIZED}), or LS (\textit{UNDERLINED}) estimation. The off-diagonal entries are omitted for clarity.}
    \begin{tabular}{cccccc}
        \toprule
         {} & {} & \multicolumn{4}{c}{Declared Radio ID} \\
         \cline{3-6}
         {SNR (dB)} & {True Radio ID} & {1} & {2}  & {3} & {4}\\
         \hline
         \multirow{4}{*}{15} & {1} & \textbf{87\%, \textit{83\%}, \underline{78\%}} &   &  &  \\
          {} & {2} &  & \textbf{91\%, \textit{87\%}, \underline{84\%}} &  & \\
          {} & {3} & & & \textbf{87\%, \textit{82\%}, \underline{79\%}} &  \\
          {} & {4} & & & & \textbf{94\%, \textit{86\%}, \underline{82\%}} \\
         \hline
         \multirow{4}{*}{18} & {1} & \textbf{94\%, \textit{87\%}, \underline{85\%}} & & &  \\
          {} & {2} & & \textbf{94\%, \textit{89\%}, \underline{87\%}} &  & \\
          {} & {3} & & & \textbf{91\%, \textit{85\%}, \underline{83\%}} & \\
          {} & {4} & & & & \textbf{96\%, \textit{88\%}, \underline{88\%}} \\
         \hline
         \multirow{4}{*}{21} & {1} & \textbf{97\%, \textit{89\%}, \underline{89\%}} & & &  \\
          {} & {2} & & \textbf{98\%, \textit{92\%}, \underline{90\%}} & & \\
          {} & {3} & & & \textbf{96\%, \textit{88\%}, \underline{89\%}} & \\
          {} & {4} & & & & \textbf{97\%, \textit{92\%}, \underline{92\%}} \\
         \bottomrule
    \end{tabular}
    \label{tab:nm_versus_ls_results}
\end{table*}
\subsection{Radio Classification Performance%
\label{sect:RF_DNA_Results}}%
The results, presented within this section, were created by dividing each Wi-Fi radio's set of $N_{B}$$=$$2,000$ collected waveforms into two sets: 1) a training set that is used to train the MDA/ML or GRLVQI classifier, and 2) a ``blind'' test set. The ``blind'' test set constitutes RF-DNA fingerprints that are not used for development nor validation of the classifier, but are used to assess its performance. All results presented here are based upon the classification of the blind test set fingerprints. Each of these waveform sets were formed using random selection.

Unlike the work in~\cite{Fadul_WCNC_2019}, both the training and test signals sets underwent Rayleigh fading and AWGN channel conditions using the approach described in Sect.~\ref{sect:Multipath_Model}. Following generation of the received signal~\eqref{eqn:rcvd_signal}, channel estimation, Sect.~\ref{sect:CH_Estimation}, and equalization, Sect.~\ref{sect:CH_EQ}, are performed prior to generation of the RF-DNA fingerprint. As in Sect.~\ref{sect:cmp_LS_NM}, a total of $N_{z}$$ =$$10$ independent, like-filtered AWGN noise realizations are generated at every SNR$\in$[9, 30]~dB, in steps of 3~dB, to facilitate Monte Carlo simulation and analysis. Training of the classifier is conducted using $k$-fold cross validation at every noise realization using $k$$=$$5$. At a given SNR, the trained MDA/ML or GRLVGI classifier that results in the minimum average percent error, across all $k$-folds and noise realizations, is designated as the ``best''. This ``best'' classifier is used in the classification of the RF-DNA fingerprints extracted from the ``blind'' test set of waveforms.
\subsubsection{Classification: LS, MMSE, and N-M Estimation%
\label{sect:results_class_ls_vs_nm}}%
The final analysis of the LS, MMSE, and N-M estimators is conducted using: the MDA/ML classifier, a Rayleigh fading channel comprised of $L$$=$$2$ paths, SNR$=$[15, 18, 21]~{dB}, and $|G_{\eta k}|^{2}$-based RF-DNA fingerprints. The N-M estimator used a total of $N_{p}$$=$$20$ candidate preambles to estimate the channel coefficients. Channel equalization is conducted using the MMSE approach described in Sect.~\ref{sect:CH_EQ}. For clarity, Table~\ref{tab:nm_versus_ls_results} presents only the diagonal elements from the average percent classification confusion matrices of each channel estimation approach. The \textit{italicized} and \underline{underlined} entries correspond to the MMSE and LS estimator-based classification results, respectively. The N-M estimator resulted in superior classification performance when compared to that of the MMSE and LS estimators at SNR$=$[15, 18, 21]~{dB} and across all Wi-Fi radios. The disparity between the N-M and LS estimators increases as SNR degrades, because, unlike the MMSE estimator, the LS estimator does not account for channel statistics. This leads to a degradation of the LS estimate as SNR is reduced. Based upon the results in Table~\ref{tab:nm_versus_ls_results} and those in Fig.~\ref{fig:error_results}, the N-M estimator is adopted for estimation of the channel coefficients for the remaining analyses.
\subsubsection{Classification: Magnitude versus Phase%
\label{sect:results_class_ls_vs_nm}}%
\begin{figure}[!b]
	\centering
	\subfigure[$L$$=$$2$ Rayleigh fading paths.]{\label{fig:2_ref_phase_amp}
	\includegraphics[width=0.75\columnwidth]{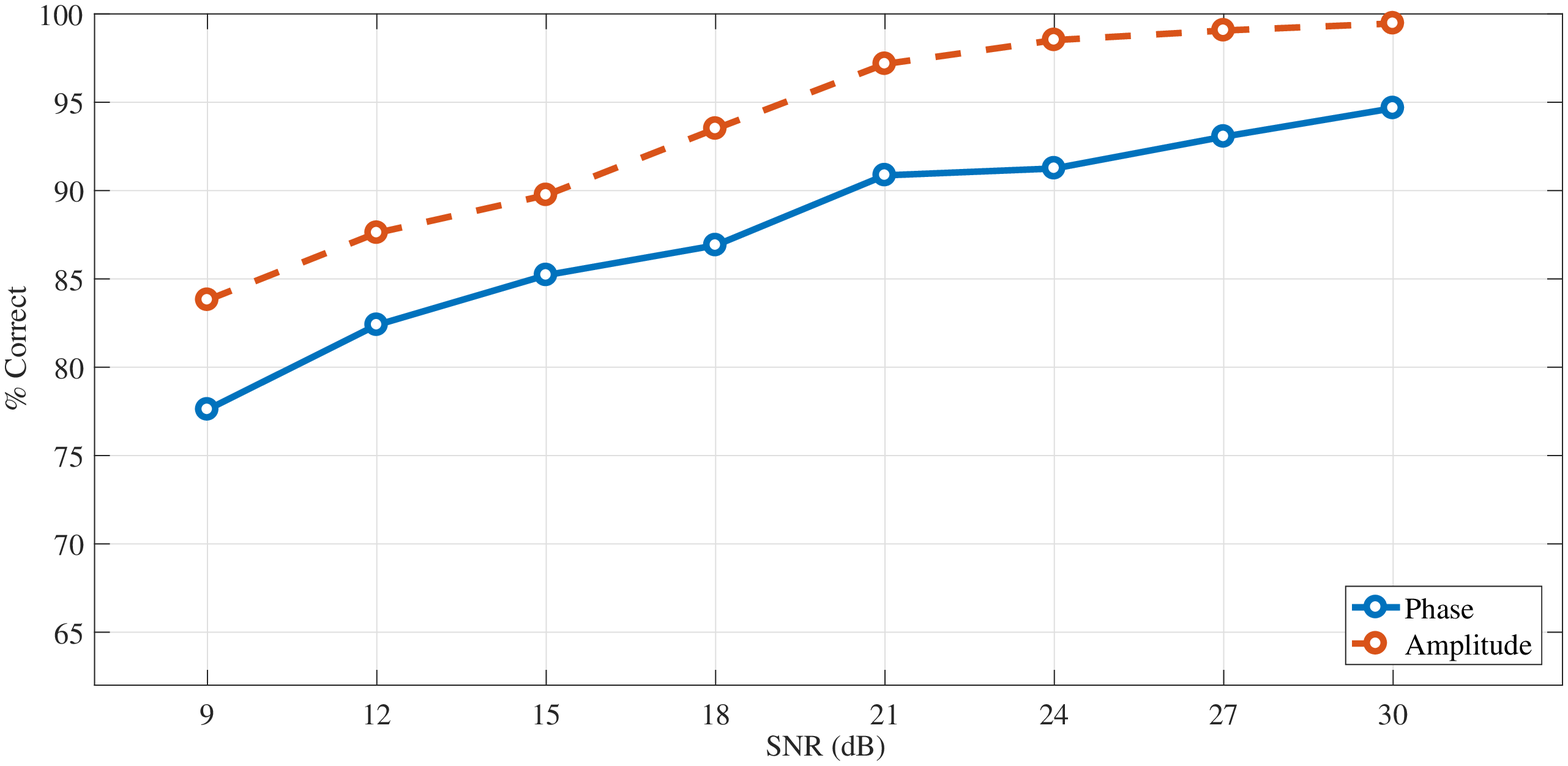}}
	\subfigure[$L$$=$$5$ Rayleigh fading paths.] {\label{fig:5_ref_phase_AMP}
	\includegraphics[width=0.75\columnwidth]{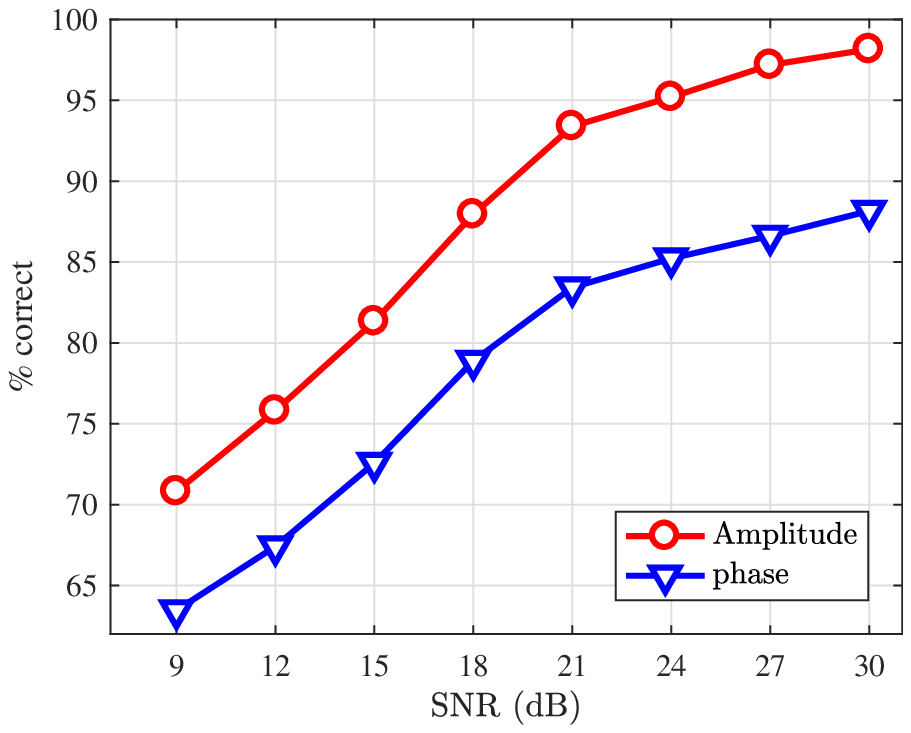}}
	\caption[]{\emph{Classification: Magnitude versus Phase:} MDA/ML percent correct classification averaged across $N_{D}$$=$$4$ devices of RF-DNA fingerprints generated from either the normalized magnitude-squared (dashed line) or phase (solid line) of the Gabor coefficients for SNR$\in$[9, 30]~dB.}
	\vspace{-5mm}\label{fig:amp_phase}	
\end{figure}

This section provides results assessing the use of RF-DNA fingerprints generated from either the normalized magnitude-squared, $|G_{\eta k}|^{2}$, or phase, $\phase{G_{\eta k}}$, response of the Gabor coefficients. The results are generated using: the MDA/ML classifier, the N-M estimator, and MMSE equalizer. The assessment is conducted using average percent correct classification performance for SNR$\in$[9, 30]~dB at 3~{dB} steps and Rayleigh fading channels consisting of $L$$=$$[2, 5]$ paths. Figure \ref{fig:amp_phase} presents percent correct classification averaged across all $N_{D}$$=$$4$ Wi-Fi radios. Classification performance is superior when using RF-DNA fingerprints extracted from the normalized magnitude-squared response of the Gabor coefficients for all SNR$\in$[9, 30]~dB. The disparity between the two RF-DNA fingerprint generation approaches becomes even greater as the number of Rayleigh fading paths increases from $L$$=$$2$, Fig.~\ref{fig:amp_phase_2}, to $L$$=$$5$, Fig.~\ref{fig:amp_phase_5}. For the $L$$=$$5$ path case, the $|G_{\eta k}|^{2}$-based RF-DNA fingerprints improves classification performance by approximately 10\% for all SNR$\in$[9, 30]~dB. Based upon the results presented here, all subsequent results are generated using RF-DNA fingerprints extracted from the normalized magnitude-squared response of the Gabor coefficients.

\subsubsection{Classification: Zero Forcing versus MMSE Equalization %
\label{sect:results_class_trad_vs_mmse}}%
This work investigates the use of an MMSE equalizer as an alternative to the ZF equalizer used in \cite{Fadul_WCNC_2019,Fadul_thesis}. The goal is to achieve the best possible RF-DNA fingerprint classification performance under degrading SNR and multipath fading conditions. As with MMSE estimation, the MMSE equalizer accounts for the channel statistics, which makes it a good equalization approach for improving RF-DNA fingerprint classification performance as SNR degrades. 
\begin{figure}[!t]
  \centering
  \includegraphics[width=0.75\columnwidth]{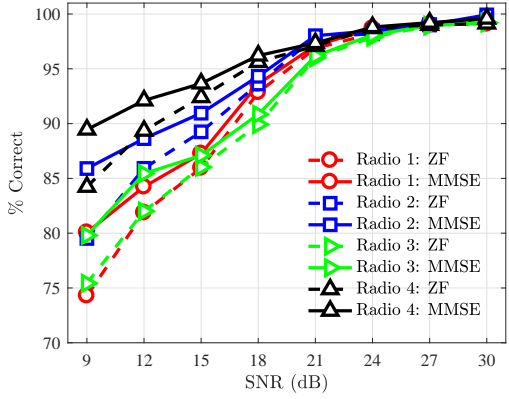}
    \caption{\textit{Classification: ZF versus MMSE Equalization:} Overlay of MDA/ML average percent correct classification performance of RF-DNA fingerprints extracted from waveforms that had undergone MMSE (solid lines) or ZF (dashed lines) channel equalization for the $N_{D}$$=$$4$ Wi-Fi radios at SNR$\in$[9, 30]~dB.}\vspace{-0.2in}
    \label{fig:results__mmse_vs_traditional}
\end{figure}
Using average percent correct classification, analysis of ZF and MMSE equalizer performance is conducted using: a $L$$=$$2$ path Rayleigh fading channel, the MDA/ML classifier, and RF-DNA fingerprints extracted from normalized magnitude-squared Gabor responses for SNR$\in$[9, 30]~dB in 3~{dB} steps. Figure~\ref{fig:results__mmse_vs_traditional} presents average percent correct classification for the $N_{D}$$=$$4$ Wi-Fi radios and both equalization techniques. For SNR$\geq$21~{dB}, there is negligible difference between the classification performance of both equalization approaches for each of the Wi-Fi radios. This is an unsurprising result as the statistics of the waveforms dominate the RF-DNA fingerprints versus that of the channel. For SNR$\leq$18~dB, the classification performance of the RF-DNA fingerprints associated with the MMSE equalizer is superior to that of the ZF equalization case. The margin between the two equalization processes actually increases as SNR degrades with the MMSE equalized case being the better of the two. At SNR$=$9~{dB}, the difference between the two approaches reaches its maximum value of approximately 5\%. This result is expected as the channel statistics become more dominant as the SNR falls. Based upon these results, the MMSE equalizer should be employed when the SNR$\leq$18~dB, while the ZF equalizer should be used for SNR$\geq$21~{dB} to reduce computational complexity.
\begin{figure*}[!t]
\begin{minipage}[b]{0.48\textwidth}
	\centering
	\subfigure[Radio \#1.]{\label{fig:Results_RFDNA_1}\vspace{2in}
	\includegraphics[width=0.85\columnwidth]{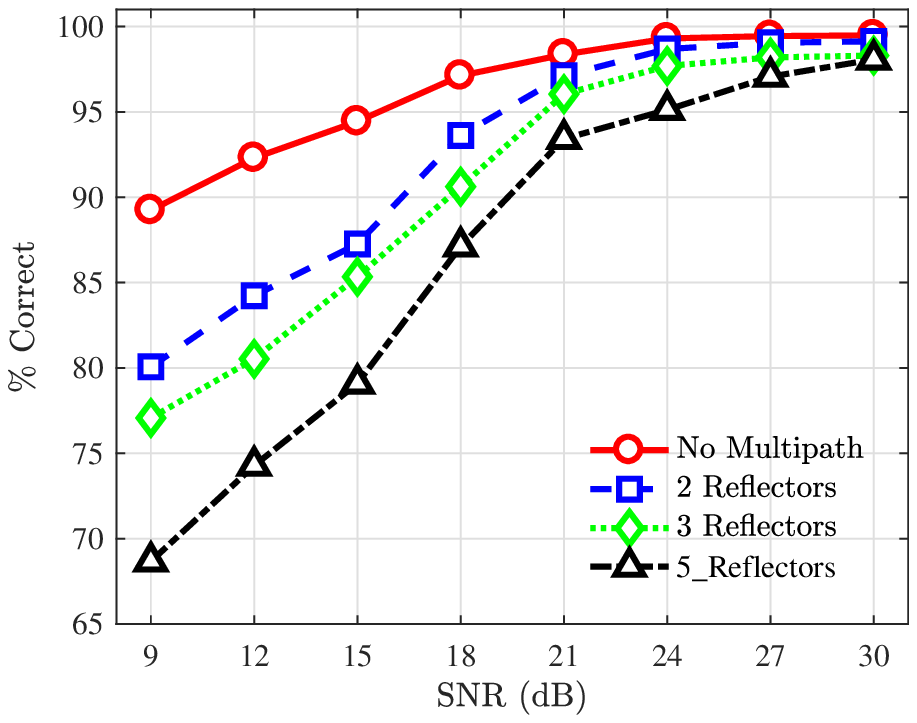}}
	\subfigure[Radio \#2.] {\label{fig:Results_RFDNA_2}
	\includegraphics[width=0.85\columnwidth]{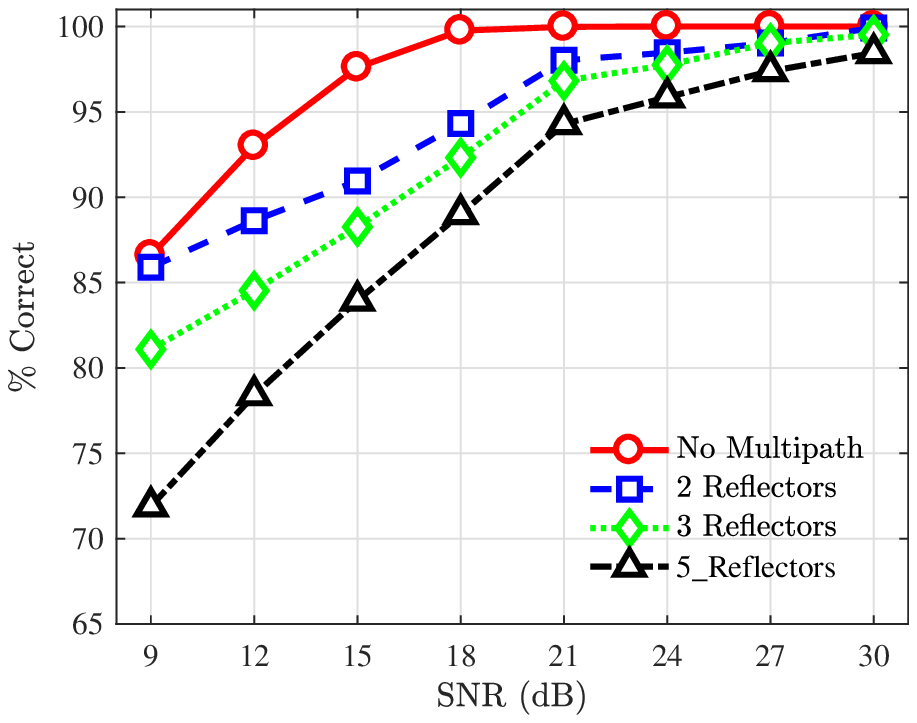}}
\end{minipage}
\begin{minipage}[b]{0.48\textwidth}
	\centering
	\subfigure[Radio \#3.] {\label{fig:Results_RFDNA_3}
	\includegraphics[width=0.85\columnwidth]{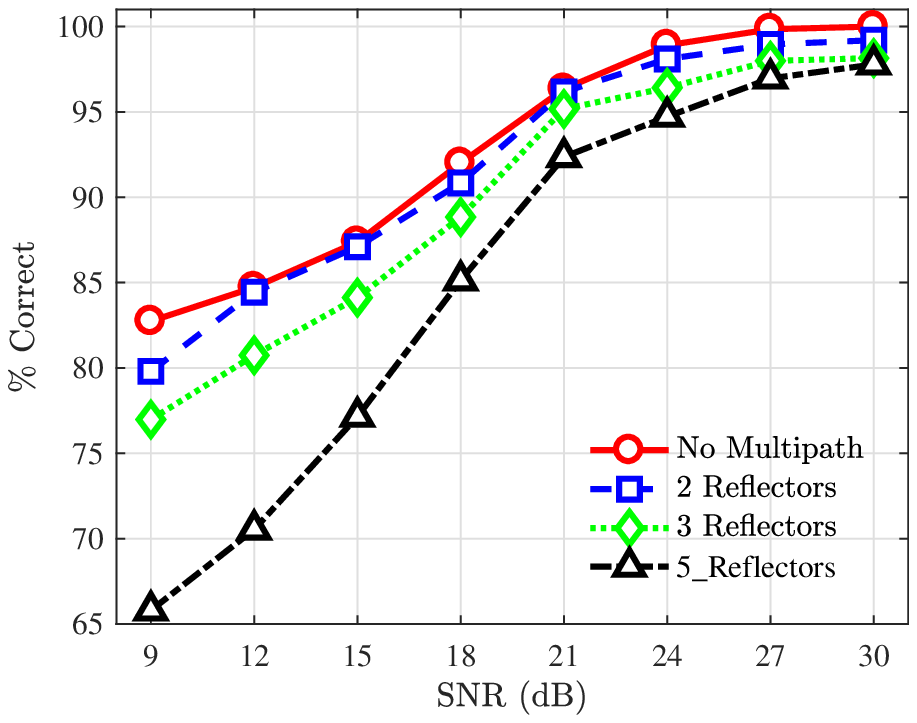}}
	\subfigure[Radio \#4.]{\label{fig:Results_RFDNA_4}
	\includegraphics[width=0.85\columnwidth]{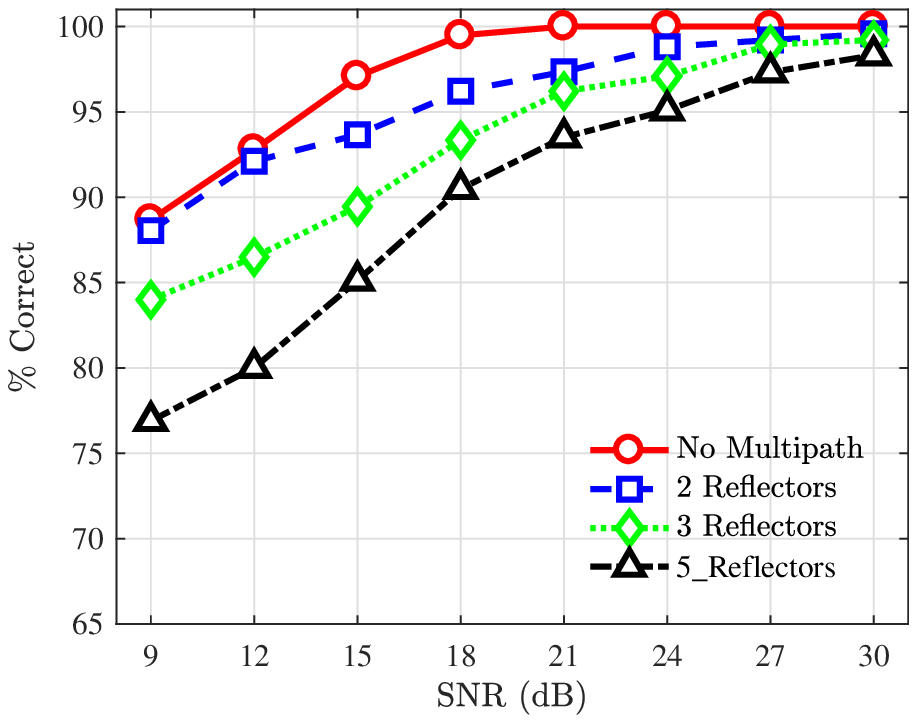}}
\end{minipage}
\caption[]{\textit{Classification: Increasing Channel Length, $L$:} MDA/ML average percent correct classification performance using RF-DNA fingerprints that were extracted from the signals of $N_{D}$$=$$4$ Wi-Fi radios that are operating under no multipath $\left(\circ\right)$ or multipath channels of length: $L$$=$$2$ $\left(\square\right)$, $L$$=$$3$  $\left(\diamond\right)$, or $L$$=$$5$  $\left(\bigtriangleup\right)$ for SNR$\in$[9, 30]~dB.}
	\vspace{-5mm}\label{fig:Fingerprinting_results}	
\end{figure*}
\subsubsection{Classification: Increasing Channel Length, $L$ %
\label{sect:results_class_increase_l}}%
This section presents average percent correct classification performance under four different multipath fading conditions: (i) no multipath, (ii) Rayleigh fading with $L$$=$$2$ paths, (iii) Rayleigh fading with $L$$=$$3$ paths, and (iv) Rayleigh fading with $L$$=$$5$ paths, Fig~\ref{fig:Fingerprinting_results}. These results are generated using N-M channel estimation, MMSE equalization, normalized magnitude-squared Gabor-based RF-DNA fingerprints, and the MDA/ML classifier for each of the $N_{D}$$=$$4$ Wi-Fi radios and SNR$\in$[9, 30]~{dB} in 3~{dB} steps. The N-M estimation is implemented using $N_{p}$$=$$20$ candidate preambles.
Overall, the best classification performance is achieved for the no multipath case across all of the Wi-Fi radios. This is attributed to the lack of residual multipath channel effects that remain after channel estimation and equalization in the other three cases. For all four Wi-Fi radios and multipath cases, average percent correct classification performance is greater than 90\% at SNR$\geq$21~dB. In comparison to the $L$$=$$2$ and $L$$=$$3$ path cases, classification results are degraded when the number of multipath paths are increased to $L$$=$$5$. This outcome is expected as the accuracy of the N-M estimated channel coefficients degrades as the number of paths increases as shown in Fig.~\ref{fig:error_results}. 

For all multipath channel conditions, the best classification performance is seen using RF-DNA fingerprints from Radio \#4, which are correctly classified at a rate of 90\% or higher for all SNR$\geq$18~{dB} as shown in Fig.~\ref{fig:Results_RFDNA_4}. The worst percent correct classification performance, across all four multipath cases, is seen in Radio \#3, Fig.~\ref{fig:Results_RFDNA_3}. Radio \#3 performance is approximately 88\% at SNR$=$18~{dB} for $L$$=$$3$, while the other three radios' RF-DNA fingerprints are classified correctly at 90\% or greater.
\begin{figure*}[!t]
\begin{minipage}[b]{0.48\textwidth}
	\centering
	\subfigure[Radio \#1.]{\label{fig:GRLVQI_1}\vspace{2in}
	\includegraphics[width=0.85\columnwidth]{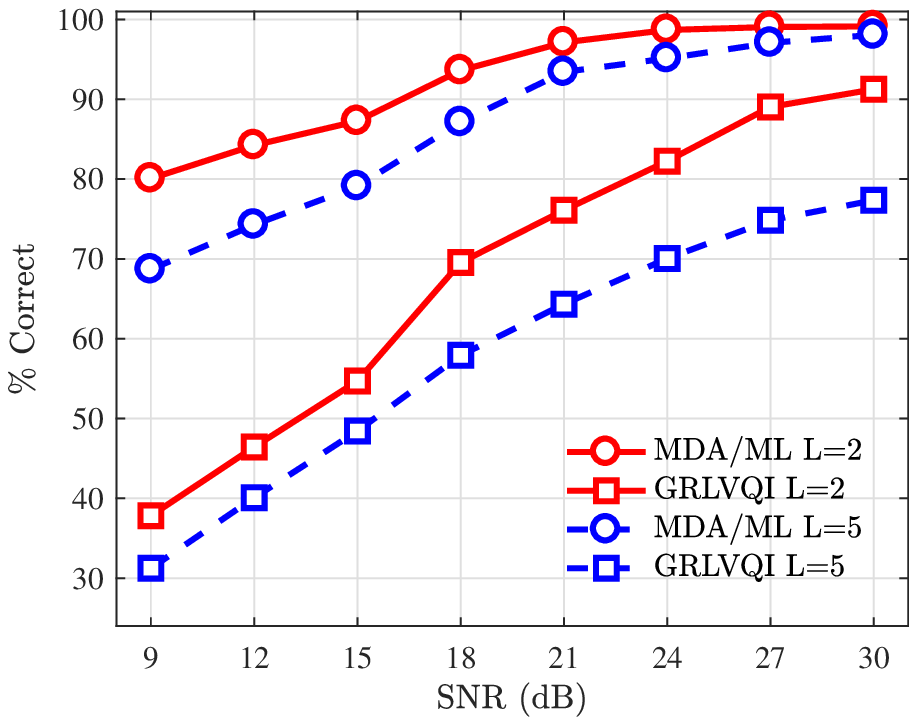}}
	\subfigure[Radio \#2.] {\label{fig:GRLVQI_2}
	\includegraphics[width=0.85\columnwidth]{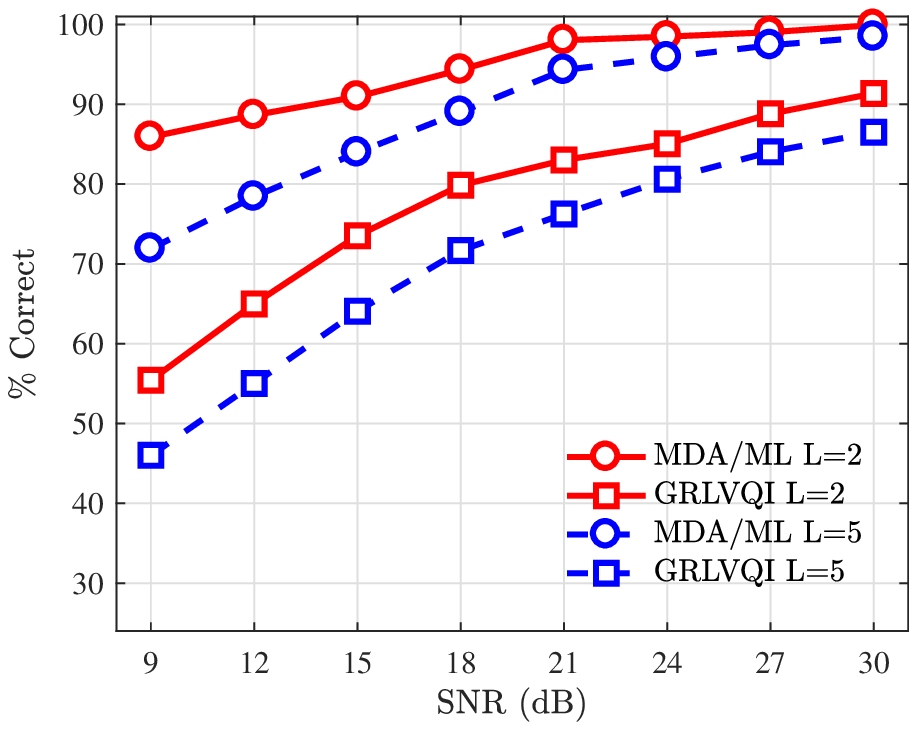}}
\end{minipage}
\begin{minipage}[b]{0.48\textwidth}
	\centering
	\subfigure[Radio \#3.] {\label{fig:GRLVQI_3}
	\includegraphics[width=0.85\columnwidth]{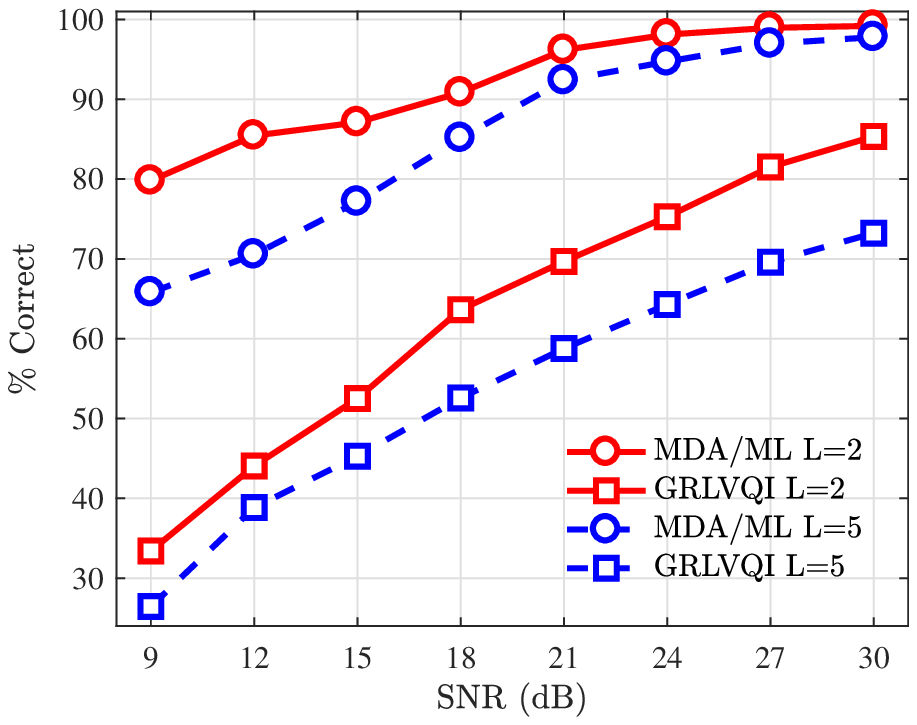}}
	\subfigure[Radio \#4.]{\label{fig:GRLVQI_4}
	\includegraphics[width=0.85\columnwidth]{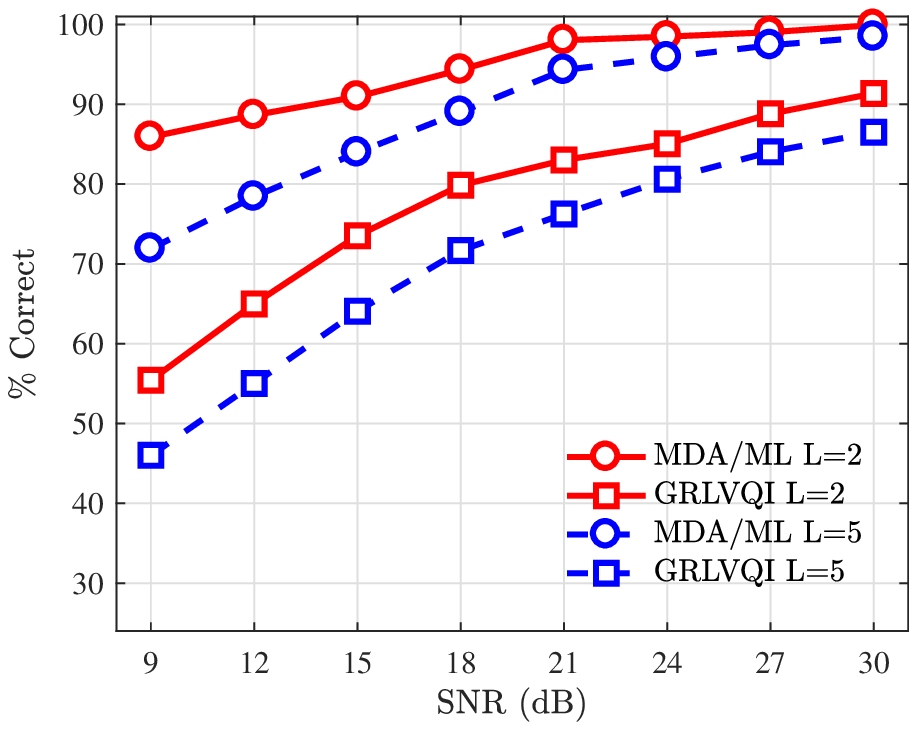}}
\end{minipage}
\caption[]{\textit{Classification: MDA/ML versus GRLVQI:} Average percent correct classification performance is presented for the MDA/ML and GRLVQI classifiers using multipath channels of length: $L$$=$$2$, or $L$$=$$5$ for SNR $\in$[9, 30]~dB.}
	\vspace{-5mm}\label{fig:GRLVQI_results}	
\end{figure*}
For the $L$$=$$2$ case, average percent correct classification performance is greater than 90\% for all four radios at SNR$\geq$18~{dB}, which is better than the results in~\cite{Fadul_WCNC_2019}. In \cite{Fadul_WCNC_2019}, average percent correct classification is greater than 90\% for all four Wi-Fi radios at SNR$\geq$21~{dB}. The 3~{dB} performance improvement is attributed to: (i) use of the MMSE equalizer and (ii) training the MDA/ML classifier using RF-DNA fingerprints extracted from waveforms that underwent multipath fading as well as N-M channel estimation and MMSE equalization. For the $L$$=$$3$ case, average percent correct classification is greater than 90\% for all four radios at SNR$\geq$21~dB. For the $L$$=$$5$ case, average percent classification is greater than 90\% at SNR$\geq$21~dB for all four Wi-Fi radios and SNR$\geq$18~{dB} for every radio except Radio \#1. Radio \#3 results in the lowest average percent correct classification performance of 66\% at SNR$=$9~{dB}.
\subsubsection{Classification: MDA/ML versus GRLVGI}%
\label{sect:results_MDAML_GRLVQI}%
This section presents the MDA/ML and GRLVQI classifier performance assessment results using N-M channel estimation, MMSE equalization, normalized magnitude-squared Gabor-based RF-DNA fingerprints, and Rayleigh fading channels consisting of $L$$=$$[2, 5]$ paths for each of the $N_{D}$$=$$4$ Wi-Fi radios and SNR$\in$[9, 30]~{dB} in 3~{dB} steps. The presented classification results were generated using the same set of RF-DNA fingerprints by both classifiers. Average percent correct classification performance per device using the two classifiers is shown in Figure \ref{fig:GRLVQI_results}. For all four Wi-Fi radios, the MDA/ML classifier outperforms that of the GRLVQI classifier for both multipath cases, $L$$=$$[2, 5]$, and across all SNRs. The disparity between the two classifiers' average percent correct classification performances actually increases as the SNR decreases. A significant difference between the two classifier approaches is that the GRLVQI requires the setting of numerous hyperparameters (e.g., learn rate, number of prototype vectors), while the MDA/ML classifier does not have hyperparameters. This is one reason that makes the MDA/ML classifier simple to implement and tractable in its use.
\section{Conclusion}%
\label{sect:SumConc}%
This work assessed the discrimination of four 802.11a Wi-Fi radios using RF-DNA fingerprints generated from the preambles of waveforms that underwent Rayleigh fading consisting of two, three, and five paths as well as subsequent channel estimation and equalization. The assessment is conducted using average percent correct classification performance over degrading channel noise for SNR$\in$[9, 30]~{dB} in 3~{dB} steps. In addition to this assessment, this work investigated: (i) N-M channel estimation performance versus LS and MMSE estimation, (ii) MMSE versus ZF equalization, (iii) classification performance using RF-DNA fingerprints extracted from either the normalized magnitude-squared or phase response of the GT coefficients, and (iv) MDA/ML versus GRLVQI classification. This work determined that the use of: N-M channel estimation, MMSE equalization, RF-DNA fingerprints generated from the normalized magnitude-squared GT coefficients, and the MDA/ML classifier resulted in the best discrimination performance of the four Wi-Fi radios. All subsequent analysis is based upon this best discrimination performance configuration.\\
\indent Average percent correct classification was greater than or equal to 90\% for all four Wi-Fi radios at SNR$\geq$18~{dB} when the Rayleigh fading channel consisted of two paths. For two paths, the classification performance never dropped below 80\% with the poorest discrimination performance being that of Radio \#3. When the Rayleigh fading channel consisted of three or five paths, the average percent correct classification was greater than or equal to 90\% for all four Wi-Fi radios at SNR$\geq$21~{dB}. Classification performance degraded the most for the five paths case with the lowest average percent correct being roughly 66\% for Radio \#3 and the highest being 77\% for Radio \#4 at SNR$=$9~{dB}. The loss in classification performance is partly attributed to the increased estimation error observed in the N-M estimator performance as the number of multipath paths changed from two, Fig.~\ref{fig:error_results_a}, to five, Fig.~\ref{fig:error_results_c}. Poorer estimation performance leads to residual channel effects, which corrupts the waveform features exploited by the RF-DNA fingerprinting process. One approach to reducing this error would be to increase the number of candidate preambles, $N_{p}$, used by the N-M channel estimator. This is under the assumption that one of the added candidate preambles provides a better representation of the received signal's coloration characteristics (i.e., reduced squared error).\\
\indent Degraded discrimination performance may also be due to the chosen features (standard deviation, variance, skewness, and kurtosis) being ill-suited due to the statistical nature of the multipath channel model. Lastly, radio discrimination was performed using the MDA/ML classifier, which makes the class assignment based upon the assumed statistical model. In this case, the statistical model was selected to be Gaussian, which is a good choice considering the use of AWGN in modeling specific SNR conditions. However, this may actually be a poor choice when multipath is present within the channel. Thus, performing RF-DNA fingerprint based radio discrimination under multipath channel conditions may be improved through the selection of: alternative, non-statistical based features, the use of a ML statistical model that better represents/fits the MDA projected RF-DNA fingerprints, a more powerful classifier, or a combination thereof.
\section*{Acknowledgment}

This work sponsored in part by the Tennessee Higher Education Commission (THEC) through the  Center of Excellence in Applied Computational Science and Engineering (CEACSE) and the University of Chattanooga (UC) Foundation Incorporated. \vspace{4mm}

\balance

\bibliographystyle{IEEEtran}

\bibliography{InfoSec2019_bib_v01} \clearpage

\end{document}